\documentclass{hep99}

\usepackage{equation,epsfig}

\begin{document}

\def\lsim{\:\raisebox{-0.5ex}{$\stackrel{\textstyle<}{\sim}$}\:}
\def\gsim{\:\raisebox{-0.5ex}{$\stackrel{\textstyle>}{\sim}$}\:}
\newcommand{\tanb}{\mbox{$\tan \! \beta$}}
\newcommand{\be}{\begin{equation}}
\newcommand{\ben}{\begin{subequations}}
\newcommand{\een}{\end{subequations}}
\newcommand{\beq}{\begin{eqalignno*}}
\newcommand{\eeq}{\end{eqalignno*}}
\newcommand{\ee}{\end{equation}}
\renewcommand{\thefootnote}{\fnsymbol{footnote}} 
\def\llgm{\left\lgroup\matrix}
\def\rrgm{\right\rgroup}

\title{Supersymmetric Higgs  pair discovery prospects at hadron colliders
\footnote{Presented by A. Belyaev at the International Europhysics Conference
on High Energy Physics, Tampere, Finland, 15-21 July 1999}
}

\author{ A       Belyaev    $^{1,2}$,  
         M       Drees      $^{3}$, 
         O  J P 'Eboli      $^{4}$, 
         J  K    Mizukoshi  $^{5}$, and 
         S  F    Novaes     $^{4}$}

%

\address{ $^1$ CERN Theory Division, CH-1211 Geneva, Switzerland\\
          $^2$ Skobeltsyn Institute for Nuclear Physics,
               Moscow State University, 119 899, Moscow, Russia\\
          $^3$ Physik Department, TU M\"unchen, James Franck Str.,
               D--85748 Garching, Germany \\
          $^4$ Instituto de F\'{\i}sica Te\'orica,
               Universidade  Estadual Paulista, \\
               Rua Pamplona 145, 01405--900, S\~ao Paulo, Brazil. \\
          $^5$ Stanford Linear Accelerator Center, University of Stanford, \\ 
               Stanford, CA 94309, USA }

\abstract{
We perform a detailed study of the potential of hadron colliders in the search for the pair
production of neutral Higgs bosons in the framework of the Minimal
Supersymmetric Standard Model.
The important role of squark loop
contributions to the signal is emphasised.  If the signal is
sufficiently enhanced by these contributions, it could even be
observable at the next run of the upgraded Tevatron collider in the
near future. At the LHC the pair production of light and heavy Higgs
bosons might be detectable simultaneously.\vspace*{-0.5cm}
} 

\maketitle

\section{Introduction}

\vskip -0.3cm
The search for Higgs bosons is one of the most important tasks for
experiments at present and future high energy colliders
\cite{higrev}. 
In particular, the Tevatron will
soon start its next collider run, with slightly increased beam energy
and greatly increased luminosity; a few years later experiments at the
LHC will commence taking data.

We study the production of two neutral Higgs bosons in
gluon fusion, followed by the decays of each boson into $b \bar{b}$
pairs. We focus on the final states where both Higgs bosons have
(nearly) the same mass, since the resulting kinematical constraint
helps to reduce the background. The SM cross section \cite{smhh} is
too small to be useful. However, the scalar sector of the SM suffers from
well-known naturalness problems. These can be cured by introducing
supersymmetry.  Here we concentrate on the simplest potentially
realistic supersymmetric model, the minimal supersymmetric standard
model (MSSM).  Several effects can greatly enhance the
Higgs pair production cross section in the MSSM with respect to  the SM:
\\
{\bf 1)}
If $\tanb \gg 1$, the Yukawa coupling of the $b$-quark is enhanced by a
factor $\sim \tanb$ with respect to its SM value. It thus becomes comparable
to the top quark Yukawa coupling for $\tanb \sim m_t(m_t)/m_b(m_t) \simeq 60$,
which is possible in most realizations of the MSSM. For Higgs boson masses
around 100 GeV the squared $b$-loop contribution then exceeds the $t$-loop
contribution, which is suppressed by the large mass of the top quark, by
a factor $\sim 15$ \cite{plehn}.
\\
{\bf 2) } For some region of parameter space ($m_A \sim
300$ GeV, $\tanb \lsim 4$) the branching ratio for $H \rightarrow hh$
decays is sizable. $h$ pair production through resonant $H$ exchange
is then enhanced by a factor $(g M_W/\lambda_t \Gamma_H)^2 \sim 100$
\cite{plehn}.
\\
{\bf 3)}   Contributions from loops involving $\tilde{b}$ or
$\tilde{t}$ squarks can exceed those from $b$ and $t$ quark loops
by more than two orders of magnitude \cite{bdemn}. This enhancement can
occur for all values of $m_A$ and \tanb, but requires a fairly light
squark mass eigenstate ($\tilde{t}_1$ or $\tilde{b}_1$), as well as
large trilinear Higgs--squark--squark couplings.

\vspace*{-0.3cm}
\section{Monte Carlo Simulation}

\vskip -0.3cm
In order to study the observability of the signal for Higgs pair
production in the $4b$ final state, we have written MC generators for
complete sets of signal as well as background processes. These
generators were designed as new external user processes for the
PYTHIA~5.7/JETSET~7.4 package~\cite{pythia}.

We used the CompHEP package \cite{comp} to generate background events
on the parton level.

For both signal and background, the effects of initial and final state
radiation, hadronization (in the string model), as well as
decay of the $b$-flavoured hadrons have been taken into account.
( see \cite{hpair-second}).

\vspace*{-0.3cm}
\section{Signal and Background Study}

\begin{figure}[htb]
\vspace*{-0.5cm}
\hspace*{-0.6cm}
\mbox{\psfig{file=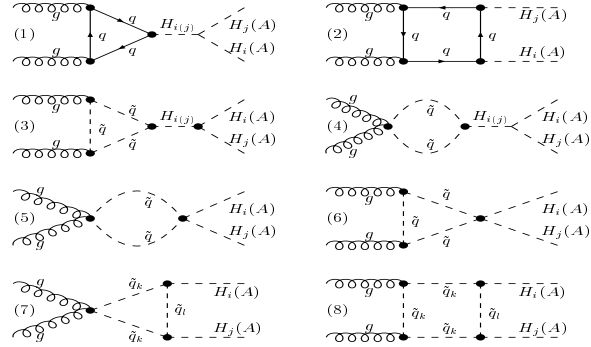,width=0.55\textwidth,height=5cm}}
\vspace*{-0.9cm}
\caption{Feynman diagrams for  $hh$, $HH$, $hH$, and $AA$ Higgs boson pair
production. $H_{i(j)} = h, H$ for $i(j)=1,2$ respectively, $\tilde
q_{k(l)} = \tilde q_1, \tilde q_2$ for $k(l)=1,2$. }
\label{fig:hh}
\end{figure}

\vskip -0.3cm
 We have calculated squark loop contributions to the
pair production of two neutral Higgs bosons. If CP is conserved,
squark loops contribute only if the two produced Higgs bosons
have identical CP quantum numbers. We gave complete
analytical expressions that allow the evaluation of these
contributions (for details  see \cite{bdemn}).
The Feynman diagrams contributing to the $g g \to hh$, $HH$,
$hH$, and $AA$ processes are presented in Fig.\ \ref{fig:hh}.
We take equal soft breaking contributions to
diagonal entries of the stop and sbottom mass matrices
($m_{\tilde{t}_L} = m_{\tilde{t}_R} = m_{\tilde{b}_R} \equiv
m_{\tilde{q}}$), as well as equal trilinear soft breaking parameters
in the stop and sbottom sectors ($A_t = A_b \equiv A_q$). We fix the
running masses of the top and bottom quarks to $m_t(m_t) = 165$ GeV
and $m_b(m_b) = 4.2$ GeV, respectively. This leaves us with a total of
5 free parameters, which determine our signal cross sections: $m_A, \
\tanb, \ m_{\tilde q},\ A_q$ and the supersymmetric higgsino mass
parameter $\mu$.

\vskip -0.2cm
This parameter space is subject to experimental constraints 
\cite{hpair-second}, especially
from the unsuccessful searches for Higgs bosons at LEP.
We also demand that the masses of the lighter physical stop and
sbottom exceed $ 90 \ {\rm GeV}$,
which follows from squark searches at LEP.
We also require that the contribution from stop and sbottom
loops to the electroweak $\rho$-parameter 
satisfies
$\delta \rho_{\tilde{t} \tilde{b}} \leq 0.0017$.
Finally, we only consider values of $A_q$ and $\mu$ in the range
$|A_q|, \ |\mu| \leq 3 m_{\tilde q}$;
this is necessary to avoid the breaking of electric charge and colour
in the absolute minimum of the scalar potential.

\begin{figure}[htb]
\vspace*{-0.6cm}
\mbox{   \epsfig{file=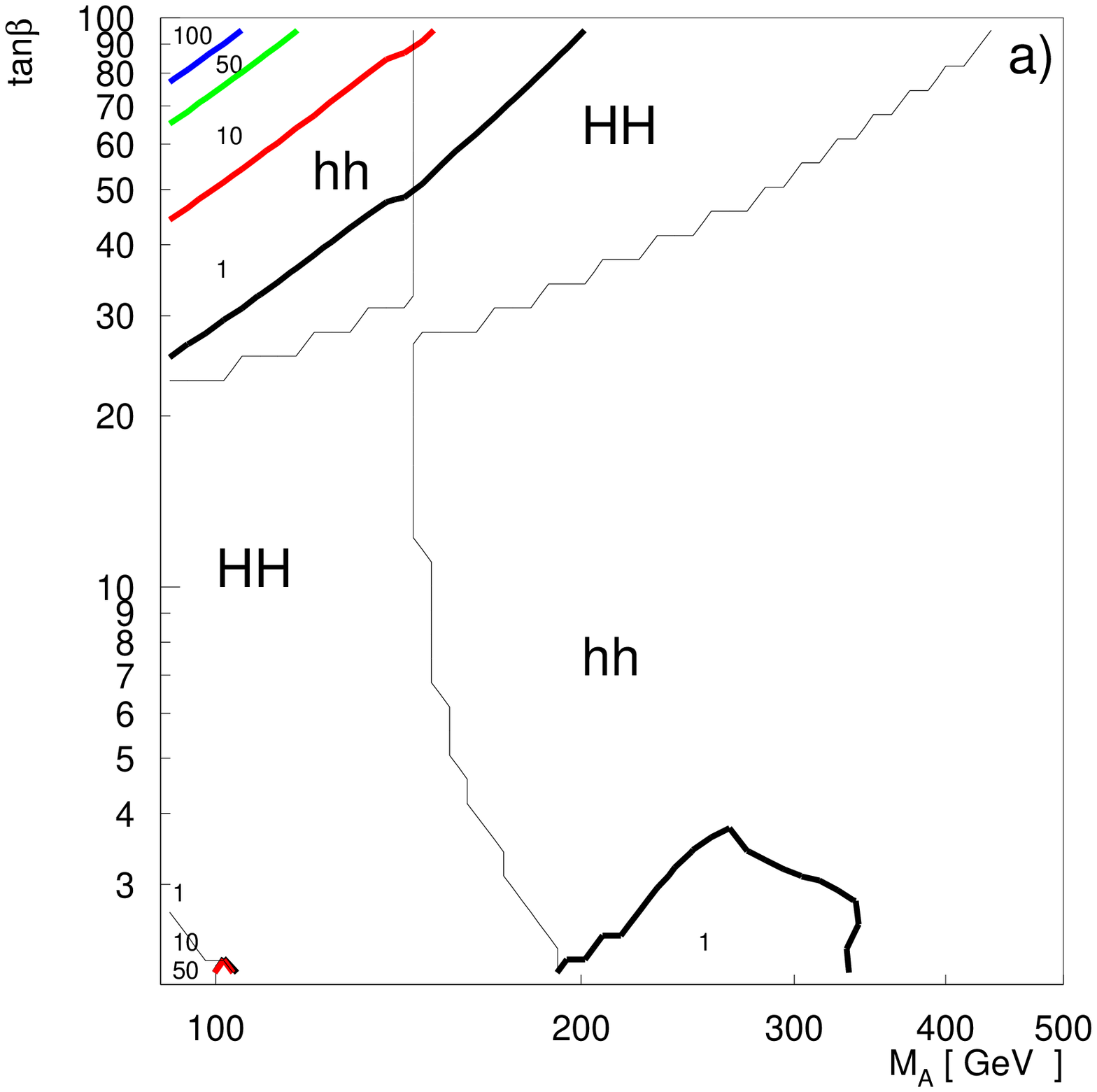,width=0.45\textwidth,height=4.5cm}}

\vspace*{-0.3cm}
\mbox{   \epsfig{file=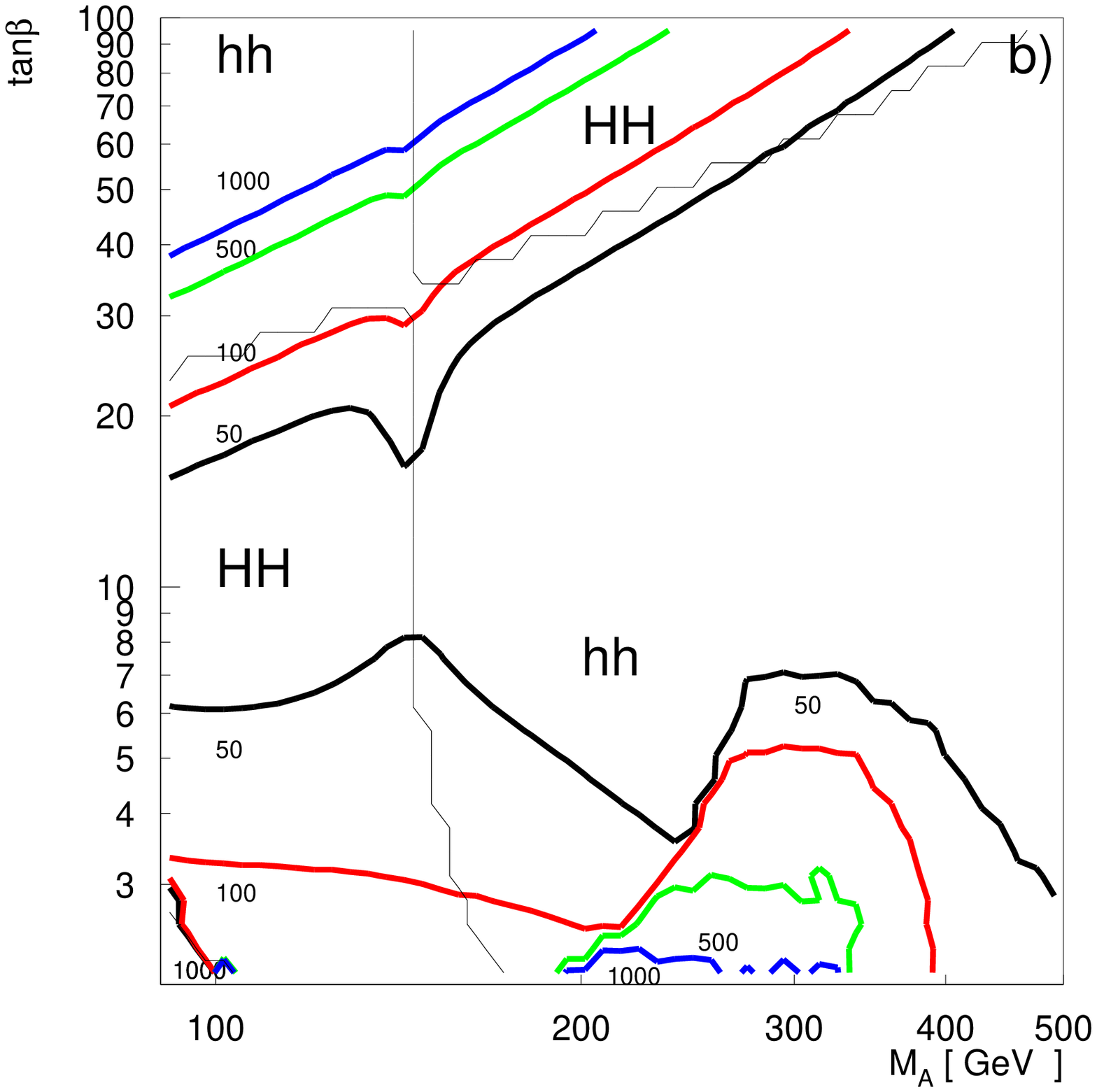,width=0.45\textwidth,height=4.5cm}}
\vspace*{-0.5cm}
\caption{Contours of constant cross section (in fb) for combined Higgs
pair production channels, for the case of negligible squark loop
contributions for the Tevatron (a) and the LHC (b).}
\label{fig:cslevel}
\end{figure}

There are 6 different channels for producing two
neutral Higgs bosons in the MSSM: $HH$, $hh$, $AA$, $Hh$, $HA$ and
$hA$. Often, several channels contribute to a given signal even after
cuts have been applied, once the experimental resolution has been
taken into account. The reason is that often two Higgs bosons are
essentially degenerate in mass, especially for high \tanb. 
In our analysis we have combined contributions from different
production channels assuming a Gaussian distribution for the
reconstructed Higgs boson mass. We start with the diagonal process
($hh, \ HH$ or $AA$ production), which  gives the best signal significance,
and then add all other contributions to the ``search window''.
 In order
to give an idea of the signal rate for negligible squark loop
contributions,  we present in Fig.~\ref{fig:cslevel} contours of
constant total signal cross section in fb in the $(m_A,\tan\beta)$
plane.

\vspace*{-0.3cm}
The total cross section is
about 200 times higher at the LHC than at the Tevatron. Given an
integrated luminosity of 100 fb$^{-1}$, we expect well over 1,000
Higgs pair events at the LHC for all combinations of $m_A$ and
\tanb. In contrast, if squark loop contributions are indeed small, at
the Tevatron the raw signal rate is often too small to give a positive
signal even at  25 fb$^{-1}$ luminocity.

In order to decide whether a Higgs pair cross section leads to a
detectable signal, we have to compute the background rate. 
To suppress ``fake'' backgrounds, we require that all four $b$-jets
be tagged as such. The total cross sections for the two main
irreducible backgrounds for the basic parton-level
acceptance cuts $p_T>25$ GeV, $\Delta R_{jj}>0.5$  for the Tevatron (LHC)
is $1.5 \;(59)$ pb for $Z b \bar{b}$ production  
and $2.6 \;(330)$ pb for $b\bar b b \bar b$ production.

The cross sections for the most important
``fake'' backgrounds for the Tevatron(LHC) are
$3.1\; (19.1) $ pb for   $Wb\bar b$ $( Q= M_{b{\bar b}})$
and $1.6 \; (164)$ nb for  $jjb\bar{b}$ $( Q= M_{b{\bar b}})$.
Since the mis-tag probability of light quark and gluon jets is
expected to be $\lsim 1\%$ \cite{d0tag,cmstag}, after $b$-tagging
these ``fake'' backgrounds are much smaller than the irreducible
backgrounds listed above and  we therefore ignore them.

One can  see that irreducible
backgrounds are clearly far larger than the signal.  
A more elaborate set of cuts is thus
necessary.

As already noted, we
require all four $b$-jets to be tagged. 
A realistic description of the $b$-tagging
efficiency is therefore very important. In case of the Tevatron, we
use the projected $b$-tagging efficiency of the upgraded D\O\ detector
\cite{d0tag} and  CMS collaboration \cite{cmstag}.
We assume that $b$-jets can be tagged only for pseudorapidity
$|\eta_b| \leq 2$ by both Tevatron and LHC experiments.

We constructed the following kinematical
variables and respective set of cuts for an efficient extraction of
the signal:
\\
{\bf 1)}
 Reconstructed Higgs boson mass, $M_H$:
we chose the pairing that gives the smallest difference between the
invariant masses of the two pairs: $M_H=[M_{b_1b_2}+M_{b_3b_4}]/2$. 
After resolution smearing, the distribution in $M_H$ for the signal can be described by
a Gaussian with width $\sigma \simeq \sqrt{M_H}$ (in GeV units).
The search window is defined as:
$0.9 m_{H,in} - 1.5 \sigma \leq M_H \leq 0.9 m_{H,in} + 1.5 \sigma.$
\\{\bf 2)} 
Mass difference between the  masses of the two pairs
(small for signal): $\left| M_{b_1b_2}-M_{b_3b_4} \right| \leq 2\sigma.$
\\{\bf 3)} The angles in the transverse plane between the two jets in each
pair should be large, while the two transverse opening angles therefore tend to be
correlated:\\
$\Delta\phi_{b_1,b_2},\ \Delta\phi_{b_3,b_4} > 1, \ \
\left| \Delta \phi_{b_1,b_2}-\Delta \phi_{b_3,b_4}
\right| <1.$
\\
{\bf 4)}
All four $b$-jets in the signal are fairly hard.
We applied cuts on the softest and hardest
of these jets, with transverse momenta $p_{T,min}$ and $p_{T,max}$:\\
{\small
${\rm TEV:} \  p_{T,min}  > M_H/8 +1.25\sigma; \
          p_{T,max} > M_H/8 + 2 \sigma $\\
${\rm LHC:} \ p_{T,min} > M_H/4; \ p_{T,max}> M_H/4 + 2 \sigma$
}\\
{\bf 5)} 
The $4b$ invariant mass:
the signal distribution for this variable is
concentrated around the invariant mass of the Higgs pair.
This quantity has been shown to be useful for disentangling
quark and squark loop contributions \cite{bdemn}:
$M_{4b} > 1.9 M_H - 3 \sigma.$
\\
The efficiency of these cuts applied plus 4$b$-tagging 
for several input (search) Higgs boson masses
is listed in the following table for the Tevatron and LHC.

The background efficiency
refers to the cross section defined through the basic
acceptance cuts ($p_T(b) > 25$ GeV for all four $b$ (anti-)quarks,
and jet separation $\Delta R_{jj} > 0.5$ for all jet pairs).

\begin{table}[htb]
{\small
\vskip -0.3cm
\begin{center}
\begin{tabular}{l   l   l   l  l}
\multicolumn{5}{c}{\bf TEVATRON:} \\
\hline
\hline
$m_{H,in}$ [GeV] &  & 120& 160 & 200    \\
\hline
\hline
$\epsilon_{\rm signal}$ [\%]&          &  2.10 &  2.74&  3.30\\
$\epsilon_{Zbb}$ [\%] &                & .187  & .0935& .0314\\
$\epsilon_{bbbb}$ [\%]&                & .137  & .0318& .0072\\
\hline
\hline
$bbbb+Zbb$      &  \# events           & 101   &  26.3&   6.6\\
$Zbb$          &  for                 &  11.1 &   5.5&   1.9\\
$bbbb$           &  25 fb$^{-1}$        &  89.9 &  20.8&   4.7\\
\hline
signal [fb] $\cdot Br$& 95\% c.l. & 37.4  & 14.6 &  7.1\\
signal [fb] $\cdot Br$& $5\sigma$ & 76.9  & 30.0 & 19.4\\
\hline
\hline
\multicolumn{5}{c}{\bf LHC:}\\
\hline
\hline
$m_{H,in}$ [GeV] &  & 120&  160   & 200         \\
\hline
\hline
$\epsilon_{\rm signal}$ [\%]&           &   .34  &   .90  &  1.38\\
$\epsilon_{Zbb}$ [\%]  &$4b$ tag        & .0263  & .0190  & .0081\\
$\epsilon_{bbbb}$ [\%]&                 & .0142  & .0112  & .0071\\
\hline
\hline
$bbbb+Zbb$ & \# events                  & 4900   & 3863   & 2419\\
$Zbb$     & for                        & 240    & 174    &  73.6\\
$bbbb$      & 100 fb$^{-1}$              &4660    &3689    &2345\\
\hline
signal [fb] $\cdot Br$& 95\% c.l.  & 570     & 171    & 70.1\\
signal [fb] $\cdot Br$& $5\sigma$  &1426     & 427    & 175\\
\hline
\hline
\end{tabular}
\end{center}
}
\vskip -0.3cm
\caption{\label{tab:eff} Signal and background efficiencies and 
minimal cross sections for a 95\% c.l. exclusion limit on, as well as a $5
\sigma$ discovery of, Higgs boson pair production at the Tevatron and LHC.} 
\vspace*{-0.5cm}
\end{table}

This table also contains results for the minimal total signal cross
section times branching ratio needed to exclude Higgs boson pair
production at the 95\% c.l., as well as the minimal total cross
section times branching ratio required to claim a $5 \sigma$ discovery
of Higgs boson pair production in the $4b$ final state. We give these
critical cross sections for two values of the integrated luminosity at
the Tevatron, characteristic for the upcoming Run II and for the final
luminosity at the end of the TeV33 run, respectively. 

Systematic uncertainties are a concern,
especially at the LHC, where the large signal rate can lead to a very
small signal-to-background ratio if the significance is defined using
statistical errors only. We  assign a 
systematic uncertainty of 2\% on the background estimate, as obtained
by extrapolation from the side bins. We thus require a minimal 
signal-to-background ratio of 0.04 for the 95\% c.l. exclusion limit, and
0.1 for the $5 \sigma$ discovery cross section. This requirement in
fact fixes the critical cross sections at the LHC for $m_{H,in} \leq
180$ GeV.

\vspace*{-0.3cm}
\section{Potential of Hadron Colliders for Higgs Pair Search}

\vskip -0.3cm
By comparing the results
of Table~\ref{tab:eff} and Fig.~\ref{fig:cslevel}a, it  becomes
clear that in the absence of sizable squark loop contributions to the
signal cross section, the potential of Tevatron experiments for this
search is essentially nil. In contrast, some parts of the $(m_A,
\tanb)$ plane can be covered at the LHC even if squark loop
contributions are negligible. For this pessimistic assumption of negligible
squark loop contributions, LHC experiments might discover a $5 \sigma$
signal if \tanb\ is large ($\gsim 50$), and can at least exclude some
regions of parameter space where \tanb\ is small ($\lsim 2.5$).

In order to illustrate the possible importance of squark loop
contributions, we performed various Monte Carlo searches of the
three-dimensional parameter space ($m_{\tilde q}, \ A_q, \ \mu$).
We believe that our procedure should reproduce the
maximal cross section to within a factor of 2 or so.

\vspace*{-0.5cm}
\begin{figure}[htb]
\mbox{ \epsfig{file=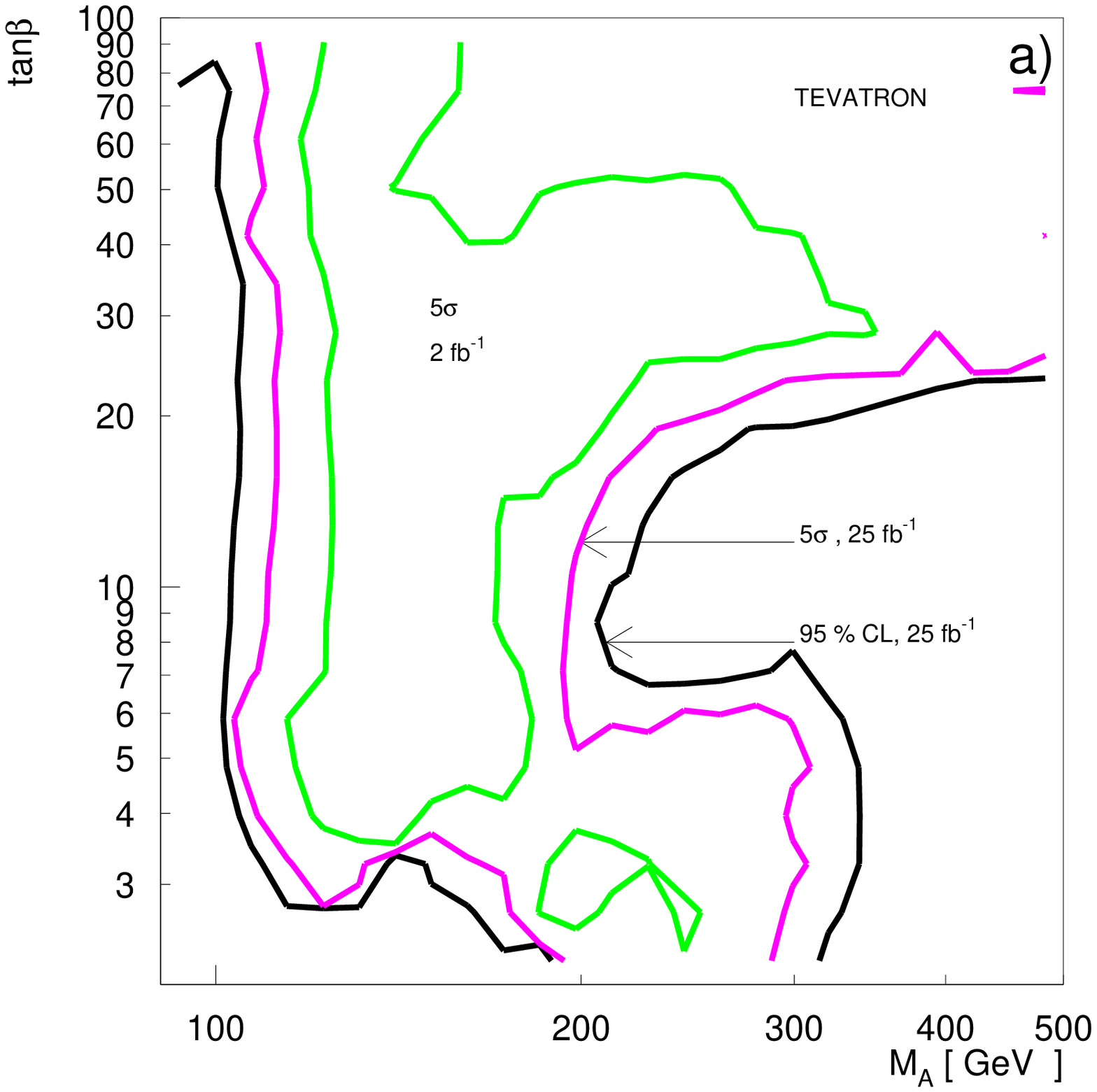,width=0.45\textwidth,height=5cm}}\\

\vspace*{-0.8cm}
\mbox{ \epsfig{file=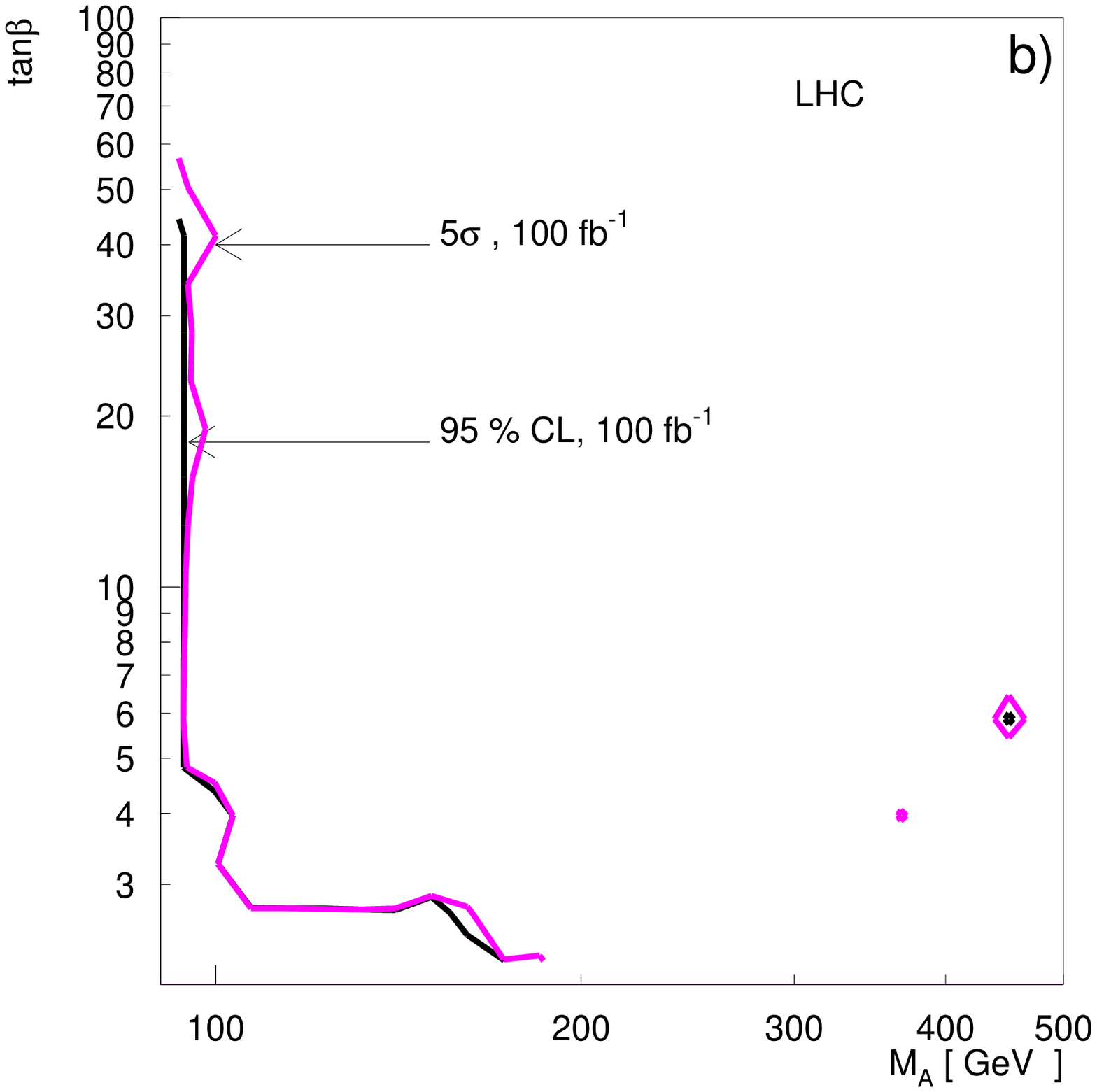,width=0.45\textwidth,height=5cm}}
\vskip -0.5cm
\caption{\label{fig:sq-lim}
95\% c.l. exclusion and $5\sigma$ discovery regions for Higgs pair
production at the Tevatron (25 fb$^{-1}$) (a) and LHC (100 fb$^{-1}$)
(b), for ``maximized'' squark loop contributions.
 The light grey contour in (a) shows the region where a $\geq
5 \sigma$ signal should be detectable at the Tevatron with just 2
fb$^{-1}$ of data. }
\end{figure}

\vspace*{-0.4cm}
The results are presented in Fig.~\ref{fig:sq-lim}, which shows the
regions that can be probed with 2 and 25 fb$^{-1}$ of data at the
Tevatron (a), and with 100 fb$^{-1}$ of data at the LHC (b). We see that
now virtually the entire part of the $(m_A,\tanb)$ plane 
will give a $ \geq 5 \sigma$ signal at the LHC.
Moreover, the entire region $m_A \leq 200$ GeV, and most of the region
with $m_A \leq 300$ GeV, can be probed at the Tevatron with 25 fb$^{-1}$
of data. Perhaps the most surprising, and encouraging, result is that a
substantial region of parameter space will give a $\geq 5 \sigma$ signal
at the Tevatron already with 2 fb$^{-1}$ of data! This is the first time
that such a robust signal for Higgs boson production at the next run of
the Tevatron collider has been suggested.

\vspace*{-0.3cm}
\section{Summary and Conclusions}

\vskip -0.3cm
The main outcome of this analysis consists in
values of the minimal total signal cross section times branching ratio
required for a 5$\sigma$ observation of the signal, as well as for
placing 95\% c.l. exclusion limits, at both the Tevatron and the LHC.

In the absence of substantial squark loop contributions, the prospects
for Tevatron experiments appear to be dim. 
LHC experiments can then only probe scenarios
with $m_A \lsim 300$ GeV and either very large or quite small values of
\tanb.

On the other hand, if squark loop contributions are nearly maximal,
and if it is possible to construct an efficient trigger for events
containing 4 $b$-jets with $\langle p_T \rangle \sim 50$ GeV, LHC
experiments should find a signal for $hh$ production for practically
all allowed combinations of $m_A$ and \tanb; $HH$ production
(augmented by nearly degenerate modes) should be visible for most
scenarios with $m_H \leq 2 m_t$. Moreover, with 25 fb$^{-1}$ of data,
Tevatron experiments would be sensitive to most of the region with
$m_A < 300$ GeV; if \tanb\ is large, even scenarios with $m_A > 500$
GeV might be detectable. Our most exciting result is that a
significant region of parameter space with $m_A \lsim 250$ GeV should
be accessible already at the next run of the Tevatron collider, which
is projected to collect 2 fb$^{-1}$ of data. This seems to be the most
robust signal for the production of MSSM Higgs bosons at the Tevatron
that has been suggested so far.

\vspace*{-0.3cm}
\section*{Acknowledgements}

\vskip -0.3cm
This work was
supported by Funda\c{c}\~ao de Amparo \`a Pesquisa do Estado de S\~ao
Paulo (FAPESP) and  by U.S. Department of Energy under the contract
DE-AC03-76SF00515.

\end{document}